\documentclass[aps,prd,preprintnumbers,nofootinbib,twocolumn]{revtex4}
\usepackage{bm}
\usepackage{latexsym}
\usepackage{dcolumn}
\usepackage{amsmath,amsfonts,amssymb}
\usepackage{graphicx,epsfig}
\usepackage{psfrag}
\usepackage{amsthm}

\def\be {\begin{equation}}
\def\ee {\end{equation}}
\def\bea {\begin{eqnarray}}
\def\eea {\end{eqnarray}}
\def\bc {\begin{center}}
\def\ec {\end{center}}
\def\bfg {\begin{figure}}
\def\efg {\end{figure}}
\def\bi {\begin{itemize}}
\def\ei {\end{itemize}}
\def\nn {\nonumber}

\def\la {\label}
\def\le {\left}
\def\ri {\right}
\def\pa {\partial}

\def\beq{\begin{equation}}
\def\eeq{\end{equation}}
\def\br{\begin{eqnarray}}
\def\er{\end{eqnarray}}
\newcommand{\eel}[1] {\label{#1}\end{equation}}

\newcommand{\bdm}{\begin{displaymath}}
\newcommand{\edm}{\end{displaymath}}

\begin{document}
\title{Quantum Raychaudhuri equation
}

\author{Saurya Das} \email[email: ]{saurya.das@uleth.ca}

\affiliation{Department of Physics and Astronomy,
University of Lethbridge, 4401 University Drive,
Lethbridge, Alberta T1K 3M4, Canada \\}

\begin{abstract}
We compute quantum corrections to the Raychaudhuri equation by replacing classical geodesics
with quantal (Bohmian) trajectories, and show that they prevent focusing of geodesics, and the formation of
conjugate points. We discuss implications for the Hawking-Penrose singularity theorems, and
curvature singularities.
\end{abstract}

\maketitle



\section{Introduction}

The celebrated Hawking-Penrose singularity theorems
in general relativity, which show that most reasonable
spacetimes are incomplete or singular in a certain precise
sense \cite{hp}, crucially depend on the validity of the Raychaudhuri
equation, via the existence of conjugate points that
the latter predicts \cite{raychaudhuri}. This equation, although being
quite general, nevertheless is completely classical in nature
and so are the singularity theorems. However, we
know that classical mechanics is an approximation of an
underlying quantum world characterized by measurement
uncertainties and the absence of particle trajectories.
As a result, when the trajectories, or geodesics,
and their congruences in the Raychaudhuri equation are
replaced with less classical/more quantum entities, one
expects corrections to the equation, and a modification
of its consequences. In particular, one may hope quantum
effects to smooth the sharp focusing of geodesics,
the formation of conjugate points and caustics, and ultimately
the spacetime singularities.

In this article, we show that by replacing the classical
velocity field used in the Raychaudhuri equation, by a
quantum velocity field,
a first order guiding equation, and
the additional {\it quantum potential}
that comes into play, this
focusing is indeed prevented.
We discuss its implications for the singularity
theorems, and curvature singularities.
Throughout, we assume a fixed (classical) background spacetime,
a four-dimensional differentiable manifold with Lorentzian
signature $(+,-,-,-)$.

Starting from a congruence of timelike geodesics
(for simplicity, our results easily generalize to null
geodesics as well) with tangent vector (`the velocity
field') $u^a(x)$ parametrized by an affine parameter $\lambda$
along the geodesics, and by $s$ for neighboring geodesics
with the deviation vector (or Jacobi field) $\eta^a$ (connecting
neighboring geodesics), it is straightforward to compute
the derivative of $u_{a;b}$ along a geodesic, as follows:
\bea
&& \frac{du_{a;b}}{d\lambda} = u_{a;b;c}~u^c = \le[ u_{a;c;b} + R_{cba}^{~~~d} u_a \ri] u^c \nn \\
&& = \le( u_{a;c} u^c \ri)_{;b} - u^c_{~;b} u_{a;c} + R_{cba}^{~~~d} u^c u_d \nn \\
&& = -u^c_{~;b}u_{a;c} + R_{cbad}~u^c u^d ~.
\la{re1}
\eea
$R_{abcd}$ and $R_{ab}$ are the Riemann and Ricci tensors respectively,
and we have used the geodesic equation $u_{a;c} u^c = 0$,
to arrive at the last line. As usual, defining the three-metric
$h_{ab} = g_{ab} - u_a u_b$, and $\theta = h^{ab} u_{a;b}$ (trace),
$\omega_{ab} = u_{[a;b]}$ (antisymmetric part),
$\sigma_{ab} = u_{(a;b)} - \frac{1}{3} h_{ab} \theta$ (traceless
symmetric part) such that $u_{a;b}=\frac{1}{3} \theta h_{ab} + \sigma_{ab} + \omega_{ab}$, the
trace of Eq.(\ref{re1}) yields the Raychaudhuri equation
\bea
\frac{d\theta}{d\lambda} = - \frac{1}{3}\theta^2 - \sigma_{ab} \sigma^{ab} + \omega_{ab}\omega^{ab}
- R_{cd} u^c u^d~. \la{re2}
\eea
For hypersurface orthogonal geodesics (i.e. $\omega_{ab} = 0$),
and when the strong energy condition via the Einstein
equations, $R_{cd} u^c u^d > 0$, is satisfied, the rhs of Eq.(\ref{re2}) is
negative, and it follows that if the congruence is initially
converging [$\theta_0 \equiv \theta(0) < 0$], the geodesics will focus,
and a caustic will develop within finite value of the affine
parameter, $\lambda \leq 3/|\theta_0|$.

To better illustrate certain points, we first take the
nonrelativistic limit of Eq.(\ref{re2}) by replacing $\lambda \rightarrow t$
(the coordinate time), $u^a(x) \rightarrow v^a(\vec x, t), (a = 1,2,3), u^0 = 1$,
and $R_{cd}u^c u^d \rightarrow \nabla^2 V$ , where $V (\vec x, t)$ is the Newtonian
gravitational potential, to obtain
\footnote{
The Raychaudhuri equation in the context of Newtonian gravity has
also been discussed in \cite{weatherall}.}
\bea
\frac{d\theta}{dt} = -\frac{1}{3} \theta^2 -\sigma_{ab}\sigma^{ab} +\omega_{ab} \omega^{ab}
-\nabla^2 V~. \la{re3}
\eea
Using the Poisson equation $\nabla^2 V = 4\pi G\rho \geq 0$,
it is easy to show that analogous focusing of particle trajectories
takes place for Eq.(\ref{re3}). Akin to using the geodesic equation
to derive Eq.(\ref{re2}), in the above, we used Newton's
second law for each particle following the flow of the velocity
field $\vec v(\vec x, t)$,
\bea
\frac{d\vec v}{dt} = -\vec\nabla V(\vec x,t)~.
\eea
Next, to obtain quantum corrections to the Raychaudhuri
equation, we first note that one now needs a quantum velocity
field. This is most easily obtained
by writing the wave function of a quantum fluid or condensate
as \cite{bohm}
%
\bea
\psi (\vec x, t) = {\cal R} e^{iS} , \la{psi1}
\eea
where $\psi (\vec x, t)$ is a normalizable wave function, and
${\cal R}(\vec x, t)$ and $S(\vec x, t)$ are real continuous functions
[e.g.for a central mass $M$ and test particle mass $m$, these are just the complete set of 
hydrogen atom wave functions, with negative (bound states) or positive (scattering states) energies, with $e^2/4\pi \epsilon_0 \rightarrow GMm$, and unitary time evolution preserving the normalization of these solutions or a superposition thereof \cite{cohen,holland}].
One adopts a statistical interpretation with $\rho \equiv |\psi|^2$
identified with the density of particles in the fluid
(the dynamics guarantees that this relation is preserved in time)
and its velocity field as
\bea
\vec v~(\vec x, t) = \frac{d\vec x}{dt}
\equiv \frac{\hbar}{m}~{\cal I}m \le( \frac{\vec\nabla\psi}{\psi} \ri) =
\frac{\hbar}{m}~\vec\nabla S (\vec x,t)~. \la{qvel}
\eea
Note that this velocity field is irrotational, $\vec \nabla \times \vec v = 0$,
i.e. $\omega_{ab} = 0$, unless $S$ is singular, signifying the presence of
vortices.
As usual, one assumes the wave function, and consequently $\vec v$ as single valued
\cite{holland}.
Substituting in the complex Schr\"odinger equation
yields two real equations
\bea
&& \frac{\pa \rho}{\pa t} + \vec\nabla \cdot \le( \rho \vec v \ri) = 0~, \la{bohmse1} \\
&& m \frac{d\vec v}{dt} = -m\vec\nabla V + \frac{\hbar^2}{2m}
\vec \nabla \le( \frac{1}{\cal R} \nabla^2 {\cal R} \ri)~. \la{bohmse2}
\eea
While Eq.(\ref{bohmse1}) is simply the probability conservation law,
Eq.(\ref{bohmse2}) resembles the classical Newton's second law of
motion but with an extra {\it quantum potential},
$V_Q\equiv - \frac{\hbar^2}{2m} \le( \frac{1}{\cal R} \nabla^2 {\cal R} \ri)$, 
in addition to the external (classical)
potential $V$ , which could be gravitational, for example.
Clearly this vanishes in the $\hbar\rightarrow 0$ limit recovering the
classical equations of motion, and all related classical predictions.
Furthermore, although Eqs.(\ref{bohmse1}) and (\ref{bohmse2}) are completely
equivalent to the Schr\"odinger equation, they can be interpreted
as giving rise to actual trajectories of particles ({\it ``quantal trajectories''})
initially distributed according to the density $|\psi|^2$, and in
quantum equilibrium, subject to the external potential
$V (\vec x, t)$, {\it as well} as the additional quantum potential $V_Q$.
Indeed the latter reproduces the observed interference
patterns in a double slit experiment, the Aharonov-Bohm effect,
the Stern-Gerlach-type experiments, and all other observed quantum phenomena,
and so long as quantum mechanics is valid, no experiments or observations
can invalidate the above picture \cite{bohm}.
Thus, we replace classical geodesics with these quantal (Bohmian) trajectories.
The Raychaudhuri equation can be rederived
for this velocity field, but now with the extra potential,
i.e. $V \rightarrow V + V_Q/m$, resulting in
\bea
\frac{d\theta}{dt} = -\frac{1}{3}~\theta^2 -\sigma_{ab}\sigma^{ab}
-\nabla^2 V + \frac{\hbar^2}{2m^2} \nabla^2 \le( \frac{1}{\cal R} \nabla^2 {\cal R} \ri)
 \la{qre1}
\eea
Now the expression for the quantum velocity field in Eq.(\ref{qvel}),
and equivalently, the presence of the quantum potential terms in Eqs.(\ref{bohmse2})
and (\ref{qre1}) ensure that the corresponding trajectories
do not cross, and there is no focusing for any value of $t$.
The easiest way to see this simple yet important result is
to note that Eq.(\ref{qvel}) is first order in time, and at any time $t$,
its right-hand side, and therefore the velocity field
are uniquely defined at each point in space. Therefore
it gives rise to nonintersecting integral curves or streamlines
\cite{bohm,deckert,figalli}.
In the above, $\psi$ itself evolves according to the Schr\"odinger equation.
Analytical as well as numerical studies indeed
demonstrate such interaction between wave packets, between
a wave packet and a barrier etc, at short distances,
and that although they can come close to each other, they
never actually meet or cross \cite{bohm,deckert,figalli,oriols}.
One may think of this as an effective repulsion between trajectories at
short distances, due to the quantum potential.
The latter of course vanishes, and the nonrelativistic Raychaudhuri equation
(\ref{re3}) is recovered in the $\hbar \rightarrow 0$ limit.

%
%
%

Relativistic generalization follows.
We start with a Klein-Gordon-type equation of the following form, in a {\it fixed} classical
background, with or without symmetries and with or without or matter
\bea
\le( \Box + \frac{m^2 c^2}{\hbar^2} -\epsilon_1 R -\epsilon_2 \frac{i}{2} f_{cd}\sigma^{cd} \ri)\Phi =0~,
\la{gokg}
\eea
where the $\epsilon_1 R$, $R$ being the curvature scalar, term admits
of the conformally invariant scalar field equation
($\epsilon_1 = 1/6$, and $m =0$), as well as that obtained from the
Dirac equation in curved spacetime ($\epsilon_1 = 1/4$) \cite{pollock,schrodinger}.
This term does not contradict observations for ray propagation
in curved spacetimes, since normally the $R=0$,
the Schwarzschild solution is used. The additional term
$-(i/2)f_{ab} \sigma^{ab}$, where $\sigma^{ab} = (1/2)[\gamma^a, \gamma^b]$, $\gamma^a$
being the Dirac matrices, and $f_{ab}$ an antisymmetric matrix, can
also be present in the second order equation derived from
the Dirac equation in curved spacetimes ($\epsilon_2 = 1$ for
fermions and $0$ for bosons) \cite{schrodinger}.
Once again, the wave function $\Phi$ is nomalizable and single valued, as required for
a quantum description of the system
(again, for example for some of the well-studied spacetimes with curvature
singularities, such as the Schwarzschild and Reissner-Nordstr\"om metrics,
these could be the wave functions in
\cite{hawking}, \cite{melnikov} or the ones used in \cite{tumulka} in the context of Bohmian
trajectories) and expressible as in Eq.(\ref{psi1}).
Note that here one has $\Phi$ on a {\it fixed} (nondynamical) background spacetime, and not part of
a coupled Einstein-scalar field system, in which both the field and the metric are dynamical.
Now the $4$-momentum, four-velocity field and ``coordinate velocity'' are defined
respectively as
\cite{sasaki,horton,durr}
\bea
&& k_a = \pa_a S \la{kgmom} \\
&& u_a = c \frac{dx_a}{d\lambda} = \frac{\hbar k_a}{m} \la{kg4vel} \\
&& \vec v  = \frac{d\vec x}{dt} = - c^2 \frac{\vec\nabla S}{\pa^0 S} ~~. \la{kgvel}
\eea
As before, one may replace the classical relativistic velocity field with the above,
which would correctly predict all observations.
Substituting Eq.(\ref{psi1}) in Eq.(\ref{gokg}) now yields the two equations
\bea
&& \pa^a \le( {\cal R}^2 \pa_a S \ri) = \frac{\epsilon_2}{2}~f_{cd} \sigma^{cd} {\cal R}^2  \la{kgcont} \\
&& k^2 = \frac{(mc)^2}{\hbar^2} - \epsilon_1 R +  \frac{\Box {\cal R}}{\cal R} \la{kgeom}
\eea
where again, Eq.(\ref{kgcont}) is the conservation equation, while Eq.(\ref{kgeom}) yields
the modified geodesic equation with the relativistic quantum potential term
$V_Q=\frac{\hbar^2}{m^2} \frac{\Box {\cal R}}{\cal R}$
\bea
u^b_{~;a} u^a = - \frac{\epsilon_1 \hbar^2}{m^2}~R^{;b} +
\frac{\hbar^2}{m^2} \le( \frac{\Box {\cal R}}{\cal R} \ri)^{;b}~.
\la{qgeod}
\eea
Then the quantum corrected Raychaudhuri equation takes the form
\bea
\frac{d\theta}{d\lambda} = && - \frac{1}{3}~\theta^2 - \sigma_{ab} \sigma^{ab}
- R_{cd} u^c u^d \nn \\
&& - \frac{\epsilon_1 \hbar^2}{m^2} h^{ab} R_{;a;b}
- \frac{\hbar^2}{m^2} h^{ab} \le( \frac{\Box {\cal R}}{\cal R} \ri)_{;a;b}~.
\la{qre2}
\eea
In this case too, the first order equation (\ref{kgvel}),
the uniqueness of the velocity field at each point in space,
and the quantum potential in Eqs.(\ref{qgeod}) and (\ref{qre2}) ensure that the
trajectories (geodesics) do not cross, again resulting in
no focusing, and no conjugate points,
for any finite value of the affine parameter.
Again, it can be seen that the quantum potential vanishes, and the
classical Raychaudhuri equation (\ref{re2}) is recovered in the
$\hbar \rightarrow 0$ limit.
Generalization to null geodesics and to Maxwell fields is straightforward
(the $m$ will not enter when these equations are derived for null geodesics
from first principles)
\cite{sasaki}.
Note that the exact form of the wave equation and its various
modifications are not important for the argument.
All that one needs
to assume is the existence of such a theory, and the first
order equations of the form (\ref{kgvel}),
and the no-crossing result continues to hold.

\section{Implications for singularity theorems}

Although the unboundedness of curvature scalars is often regarded as a signature of singular spacetimes, this is neither necessary (e.g. removing a wedge from Minkowski space makes it singular) nor sufficient (e.g. when they are reachable only in infinite proper time, or the difficulties in specifying singularity as a ``place'' for generic spacetimes). Therefore
one equates the incompleteness of geodesics (which is easier to determine), equivalently the termination of existence of a particle (or photon), to singular or
pathological spacetimes \cite{wald,hp}.
It can be shown that the focusing of geodesics implies the existence of pairs
of conjugate points, where $\vec \eta$ vanishes for neighboring
geodesics, which in turn implies that sufficiently long
geodesics cannot be maximal length curves. The existence
of maximal geodesics is predicted on the other hand
by a set of global arguments for globally hyperbolic
spacetimes. This apparent contradiction is resolved
by requiring that sufficiently long geodesics cannot
exist, leading to geodesic incompleteness and ``singular spacetimes'',
which is the essence of the singularity theorems (as mentioned earlier,
throughout this article, we omit the finer distinction
between timelike and null geodesics, since we expect
our results to hold for either) \cite{wald}.
However, we know that in the quantum picture,
particles {\it do not} follow classical trajectories or geodesics; therefore the
Hawking-Penrose singularity theorems, although still valid, lose much of their
original motivation, 
and therefore need to be replaced by a quantum version.
As we have shown here,
particles can be thought of following quantal (Bohmian) trajectories instead
(as they correctly predict all observations);
therefore these are natural candidates for replacing geodesics in the singularity
theorems. However, since these are complete (i.e. do not end) and
do not have conjugate points (i.e. $\vec \eta$ never vanishes),
the resultant ``semiclassical" version of the singularity theorems
now {\it do not} predict the existence of singularities.
Furthermore, as shown below, regions of unbounded curvature
are never reached by the quantal trajectories. Therefore,
either one would have to find another way to characterize
singularities using quantum mechanics,
and applicable to a wide class of spacetimes,
or would have to conclude that singularities are in fact avoidable.

\section{Implications for curvature singularities}

Next, consider the geodesic deviation equation modified
by the quantum potential term (we omit the $\epsilon_1$ term here)
\bea
\frac{D^2 \eta^a}{d\lambda^2} = -\frac{1}{c^2}  R^a_{~bfc} u^b u^c \eta^f
%
- \frac{\hbar^2}{m^2 c^2} \le[\le( \frac{\Box {\cal R}}{\cal R} \ri)^{;a} \ri]_{;c} \eta^c~.
\la{qgde}
\eea
For spacetimes in which curvature scalars
(such as the Kretschmann scalar $R_{abcd} R^{abcd}$) blow up
(e.g. at $r=0$ for certain black holes), the deviation
vector $\vec\eta\rightarrow 0$. Since the quantum Raychaudhuri
equation, as well as the additional term in Eq.(\ref{qgde}) above
show on the other hand that $\vec\eta \neq 0$ at all times,
these extreme curvature regions are not accessible, and
observed curvature
components and scalars would also remain finite (albeit
large) at all times.

To summarize, we have shown that replacing classical
trajectories or geodesics by their quantum counterparts
gives corrections to the Raychaudhuri equation, which naturally
prevents focussing and the formation of conjugate points.
Therefore, if one replaces classical geodesics with quantal (Bohmian)
trajectories in the singularity theorems, then the quantum version
of these theorems do not show that spacetime singularities are inevitable.
%
We reiterate that we have simply rewritten regular quantum mechanics
in a convenient form, in which the no crossing of trajectories due to the first order
evolution equation becomes transparent.
Another way of looking at this is that the quantum potential, although being small,
causes deviations from classical trajectories at short distances,
sufficient for trajectories to not cross each other.
Also, we have not assumed spherical or any other symmetry
in our analysis, and our results are valid for all spacetimes.
Our results hold for bosons as well as fermions
(note that we have included the Dirac equation),
although for fermions, one
might encounter additional {\it exchange forces}
at small distances, further inhibiting the focusing of geodesics.
To our knowledge, this is the first time that systematic quantum corrections to
the Raychaudhuri equation have been computed and its implications examined, without
using any specific formulation of quantum gravity, or invoking special symmetries.
It is tempting to speculate that for curved spacetimes, the
quantum potential becomes important, and the no convergence
would be seen near the Planck
length, the latter being the natural scale in quantum
gravity. It would be interesting to investigate the
fate of these quantum trajectories for values of the affine
parameter near or exceeding $3/|\theta_0|$. A combination of
analytical and numerical studies should shed more light
on these issues.
Finally, it may be argued that our assumption of a smooth background
manifold may break down at small scales, and especially in regions of high
curvatures, being replaced by a more fundamental ``quantum structure''. This is certainly a
possibility, although perhaps not compelling. Furthermore, as remarked earlier, and
as our Eq.(\ref{qgde}) suggests, regions of very high curvatures may in fact be
inaccessible.

\begin{center}
{\bf Acknowledgments}
\end{center} 

I thank
S. Braunstein, R. K. Bhaduri, A. Figalli, D. Hobill, S. Kar,
G. Kunstatter, R. B. Mann, R. Parwani, T. Sarkar, L. Smolin and R. Sorkin
for discussions and correspondence, and also the anonymous referees for useful
suggestions which helped improve the manuscript,
and the IQST and PIMS, University of Calgary, for hospitality, where part of this work
was done. I also thank the Perimeter Institute for Theoretical Physics for
hospitality through their affiliate program.
This work is supported by the Natural Sciences and Engineering Research Council of Canada.



\end{document}